\documentclass[sigconf]{acmart}

\AtBeginDocument{%
  }

\usepackage{enumitem}
\usepackage{tcolorbox}
\usepackage[frozencache,cachedir=.]{minted}
\usepackage{soul}
\usepackage{xspace}
\usepackage{multirow}
\usepackage{xurl}
\usepackage{url}
\usepackage{pifont}
\usepackage{hyperref}
\usepackage{colortbl}
\usepackage{adjustbox}
\usepackage{algorithm}
\usepackage{algpseudocode}
\usepackage{amsmath}
\usepackage{listings}
\AtBeginDocument{\DeclareCaptionSubType{lstlisting}}
\usepackage{subcaption}
\usepackage{booktabs}
\usepackage{subfloat}

\usepackage{amssymb}
\usepackage{tikz}
\usepackage{setspace}

\definecolor{dawnblue}{rgb}{0.84, 0.92, 1.0}

\definecolor{LightGray}{gray}{0.9}
\newcommand{\tool}{\textsf{LOVA}\xspace}
\newcommand{\toolc}{\textsf{LOVA-C}\xspace}
\newcommand{\toola}{\textsf{LOVA-A}\xspace}
\newcommand{\toolv}{\textsf{LOVA-V}\xspace}

\newcommand{\cvefixes}{\textit{CVEFixes}\xspace}
\newcommand{\cvefixesc}{\textit{CVEFixes-C}\xspace}
\newcommand{\cvefixesp}{\textit{CVEFixes-P}\xspace}
\newcommand{\cvefixesj}{\textit{CVEFixes-J}\xspace}
\newcommand{\cvefixespm}{\textit{CVEFixes-P-M}\xspace}
\newcommand{\cvefixesjm}{\textit{CVEFixes-J-M}\xspace}

\newcommand{\bigvul}{\textit{Big-Vul}\xspace}
\newcommand{\smartfix}{\textit{SmartFix}\xspace}

\usepackage{xcolor}
\usepackage{mdframed}

\usepackage{fontawesome}

\definecolor{cyan}{rgb}{0.0, 0.72, 0.92}
\hypersetup{
    colorlinks = true,
    linkcolor = red,
    anchorcolor = red,
    citecolor = red,
    filecolor = red,
    urlcolor = red
}

\begin{document}

\title{If LLMs Would Just Look: Simple Line-by-line Checking \\ Improves Vulnerability Localization}

\author{Yue Li}
\affiliation{%
  \institution{National Key Lab for Novel Software
\\ Technology, Nanjing University}
  \city{Nanjing, Jiangsu}
  \country{China}}
\author{Xiao Li}
\affiliation{%
  \institution{National Key Lab for Novel Software
\\ Technology, Nanjing University}
  \city{Nanjing, Jiangsu}
  \country{China}}
  \author{Hao Wu}
\affiliation{%
  \institution{National Key Lab for Novel Software
\\ Technology, Nanjing University}
  \city{Nanjing, Jiangsu}
  \country{China}}
  \author{Yue Zhang}
\affiliation{%
  \institution{Shandong University}
  \city{Qingdao, Shandong}
  \country{China}}
    \author{Xiuzhen Cheng}
\affiliation{%
  \institution{Shandong University}
  \city{Qingdao, Shandong}
  \country{China}}
    \author{Yating Liu}
\affiliation{%
  \institution{National Key Lab for Novel Software
\\ Technology, Nanjing University}
  \city{Nanjing, Jiangsu}
  \country{China}}
    \author{Fengyuan Xu}
\affiliation{%
  \institution{National Key Lab for Novel Software
\\ Technology, Nanjing University}
  \city{Nanjing, Jiangsu}
  \country{China}}
    \author{Sheng Zhong}
\affiliation{%
  \institution{National Key Lab for Novel Software
\\ Technology, Nanjing University}
  \city{Nanjing, Jiangsu}
  \country{China}}

\begin{abstract}
The rapid expansion of software systems and the growing number of reported vulnerabilities have emphasized the importance of accurately identifying vulnerable code segments. Traditional methods for vulnerability localization, such as manual code audits or rule-based tools, are often time-consuming and limited in scope, typically focusing on specific programming languages or types of vulnerabilities. 
In recent years, the introduction of large language models (LLMs) such as GPT and LLaMA has opened new possibilities for automating vulnerability localization. 
However, while LLMs show promise in this area, they face challenges, particularly in maintaining accuracy over longer code contexts. 
This paper introduces \tool, a novel framework leveraging the self-attention mechanisms inherent in LLMs to enhance vulnerability localization. 
Our key insight is that self-attention mechanisms assign varying importance to different parts of the input, making it possible to track how much attention the model focuses on specific lines of code. 
In the context of vulnerability localization, the hypothesis is that vulnerable lines of code will naturally attract higher attention weights because they have a greater influence on the model's output. 
By systematically tracking changes in attention weights and focusing on specific lines of code, \tool improves the precision of identifying vulnerable lines across various programming languages. 
Through rigorous experimentation and evaluation, we demonstrate that \tool significantly outperforms existing LLM-based approaches, achieving up to a \textit{5.3$\times$} improvement in F1-scores. \tool also demonstrated strong scalability, with up to a \textit{14.6$\times$} improvement in smart contract vulnerability localization across languages like C, Python, Java, and Solidity. Its robustness was proven through consistent performance across different LLM architectures.
\end{abstract}





\maketitle

\section{Introduction}
Software vulnerabilities are weaknesses or flaws that attackers can exploit to compromise a system’s security, functionality, or data integrity.
As of September 2024, over 240,000 vulnerabilities~\cite{cve_mitre} have been publicly reported in the Common Vulnerabilities and Exposures (CVE) database, and these vulnerabilities can have serious consequences for the security and stability of a software system. 
Accurately localizing vulnerabilities is increasingly vital and has become a key area of focus~\cite{abreuAccuracySpectrumbasedFault2007, moonAskMutantsMutating2014, liDeepFLIntegratingMultiple2019} in recent years. 
It allows developers to isolate the specific section or module of code where the flaw exists, enabling targeted and efficient remediation. 
This precision minimizes the risk of unnecessary changes that can introduce new issues, significantly reducing the time and resources required to fix vulnerabilities. 
Without precise localization, the process of addressing vulnerabilities can become inefficient, leading to potential system-wide disruptions and further security risks.

Large language models (LLMs), such as GPT~\cite{achiam2023gpt} and LLaMA~\cite{touvron2023llama}, are sophisticated AI systems designed to process vast amounts of data to understand and generate text or code with human-like accuracy. 
These models have become prominent due to their advanced capabilities in tasks such as natural language understanding, translation, summarization, and, more recently, in the domain of software development and security. 
One emerging use case is vulnerability localization, where LLMs help identify and isolate security flaws within large and complex codebases. 
Traditionally, vulnerability localization has been a time-consuming and labor-intensive process, often requiring manual code audits or the use of rule-based tools~\cite{wongSurveySoftwareFault2016b, zouEmpiricalStudyFault2021}.  
These methods not only demand significant expertise but are also prone to human error, potentially leading to overlooked vulnerabilities.
Additionally, rule-based tools are often limited in scope, typically targeting specific programming languages or certain types of vulnerabilities. 
LLMs, however, offer a new level of automation and intelligence in this area. 
LLMs can analyze extensive codebases and detect patterns that are commonly associated with vulnerabilities. 
These models are trained on massive datasets that include programming languages, software repositories, and vulnerability databases.  
By pasting the code into an LLM and explicitly requesting it to localize the vulnerability, the LLM can efficiently pinpoint the location of the vulnerable code. \looseness=-1

While LLM has demonstrated potential for vulnerability localization, there are several notable limitations. 
First, LLMs struggle to maintain reasoning accuracy over longer code contexts~\cite{liu2024lost, kuratov2024babilong}. 
For instance, in our exploratory experiments, we observed a significant drop in the output accuracy as the lines of code (LoC) increased: when the LoC exceeded 300, the accuracy fell below 5\%. 
Although more advanced reasoning techniques such as Self-Consistency~\cite{wang2022self}-—which analyzes the code multiple times using different reasoning paths to ensure consistency, 
and rStar~\cite{qi2024mutual}-—which uses a self-play generation-discrimination process with Monte Carlo Tree Search (MCTS), 
where two similar models mutually verify reasoning paths—can improve accuracy, these methods still fail to effectively address the issue of long context degradation. 
Moreover, they rely on stochastic decoding strategies, which introduce randomness into the results and hinder accurate vulnerability localization.
Even methods like majority voting resulted in only marginal improvements. 

Our work aims to enhance the accuracy of vulnerability localization by leveraging self-attention mechanisms (self-attention allows the model to assign varying importance to different input elements, which is the core of LLMs). 
The reason lies in the intrinsic connection between self-attention and the model's output, making it a common tool for interpreting model behavior.
We hypothesize that if an LLM can classify vulnerabilities (e.g., types~\cite{yin2024multitask, liu2024vuldetectbench}), it should also inherently possess the capability to pinpoint their exact locations. 
This is based on the premise that code with vulnerabilities is more likely to significantly influence the model's attention mechanisms and its subsequent outputs. 
However, contemporary LLMs face challenges in accurately localizing vulnerabilities due to their processing of extensive contextual information. 
This often results in key lines or code segments being abstracted into more generalized, higher-level information, causing a loss of detail and clarity. 
Therefore, the critical task is to refine the model's focus to concentrate specifically on the lines of code that are vulnerable, enhancing the precision of vulnerability detection and localization.\looseness=-1

To address this problem,  we propose a strategy whereby, during the construction of prompts, the LLM is deliberately directed to refocus its attention on areas of the code identified as potentially vulnerable. 
However, given the inherent difficulty in pre-identifying which specific sections of code contain vulnerabilities, it is not feasible to target the LLM’s attention exclusively to these areas. 
Instead, our approach involves directing the LLM's focus across all code lines, both vulnerable and non-vulnerable. 
This is predicated on the understanding that non-vulnerable code segments have a minimal impact on the model’s overall output, allowing for comprehensive scrutiny without compromising the effectiveness of the results. 
This scenario is analogous to a student who is familiar with all the answers on an exam but commits errors due to the extensive length of the test. 
By mandating a review of each response, the student has the opportunity to identify and rectify any mistakes, thereby ensuring that the correct answers remain unaffected. \looseness=-1

Building on this idea and our exploratory experiments (refer to \S\ref{subsec:ee} for validation through a series of experiments), we developed \tool (\textit{LO}-cating \textit{V}ulnerabilities via \textit{A}ttention), with three key stages: code line highlighting, attention calculation, and vulnerability localization. 
In the code line highlighting stage, two versions of the input are generated: a baseline version and a set of highlighted versions. 
In the attention calculation stage, attention maps are produced for both versions, and the differences between them are analyzed to assess the impact of highlighting specific lines. 
Finally, in the vulnerability localization stage, the attention patterns are examined to locate vulnerabilities in specific lines.  \looseness=-1

During the development of \tool, we overcame several key challenges. \textit{First}, simply highlighting code by adding comments can be ineffective in large code contexts, leading to poor attention enhancement or even hallucinations. To resolve this, we designed a line index-based prompt that avoids adding extraneous content, thus preventing confusion for the model. \textit{Second}, analyzing large attention tensors is impractical due to the massive size of the attention outputs, which can contain millions of elements. To address this, we implemented a novel dimensionality reduction technique to aggregate attention across heads and layers, focusing on critical patterns relevant to vulnerability detection. \textit{Finally}, we employed a language-aware deep learning model to classify the reduced attention matrix, enabling more precise vulnerability identification across different programming languages.

The evaluation of \tool reveals its superiority in localizing vulnerabilities compared to traditional LLM-based approaches across different programming languages and benchmarks. In terms of effectiveness (RQ1), \tool achieved a significant improvement in F1-scores on both the \bigvul and \cvefixesc datasets, outperforming vanilla output and other sophisticated LLM techniques such as CoT, MoA, and rStar by up to \textbf{5.3$\times$}. This high performance stems from the model’s ability to balance precision and recall, minimizing false positives while maintaining high accuracy in identifying vulnerabilities. Particularly in longer contexts, \tool demonstrates resilience where other techniques falter, maintaining high Top-N accuracy across various LoC ranges. In terms of scalability (RQ2), \tool showcases its generalization capabilities by performing effectively across multiple languages, including C, Python, Java, and Solidity. It achieves up to a \textbf{14.6$\times$} improvement in contract-level vulnerability localization for smart contracts, underscoring its adaptability to diverse coding environments. The robustness of the system (RQ3) is further demonstrated in its cross-compatibility with various LLM architectures, consistently outperforming baseline methods regardless of the underlying model. Finally, the ablation study (RQ4) highlights the importance of each design component, confirming that the integration of contextual information and attention calculation mechanisms is essential to the method's overall effectiveness in vulnerability localization.

In summary, we make the following contributions.
\begin{itemize}
    \item We are the first to discover that the self-attention mechanism of LLMs can be effectively utilized for vulnerability localization.  We demonstrate that by tracking changes in attention weights, it is possible to identify specific lines of code that are likely to be vulnerable.
    \item We design and implement \tool, a novel framework for vulnerability localization. Key techniques of \tool include a line index-based prompt design to prevent confusion in large code contexts, dimensionality reduction to simplify massive attention outputs, and the use of attention map differences to assess the impact of focusing on specific lines. Additionally, a language-aware deep learning model is employed to generalize the approach across multiple programming languages, making it scalable and effective for diverse codebases.
    \item We demonstrate that \tool significantly outperforms traditional LLM-based approaches. It shows improvements in precision, recall, and scalability across multiple programming languages, such as C, Python, Java, and Solidity. Moreover, it adapts well to different code lengths and architectures, ensuring robustness.
\end{itemize}

\section{Background}
\label{sec:llm}

\subsection{LLM Architecture}
\label{subsec:llmarch}

A large language model (LLM) is an advanced type of artificial intelligence designed to understand and generate human-like text. These models are typically built using deep learning techniques, specifically neural networks with a large number of parameters, trained on vast amounts of text data.  LLM is typically based on the transformer architecture \cite{vaswani2017attention}.\looseness=-1

A transformer is a type of deep learning model architecture. The key innovation of the transformer architecture is its use of self-attention mechanisms, which allow the model to weigh the importance of different words in a sentence, regardless of their position (which will be introduced next).
The transformer architecture consists of an encoder and a decoder: the encoder processes the input sequence into context-rich representations, which the decoder then uses, along with previously generated tokens, to produce the final output sequence. Since both the transformer and its sub-components (the encoder and decoder) are capable of handling sequence-to-sequence generation tasks, and each component has different functions, the development of current LLMs has branched into three main technical routes: encoder-decoder (such as T5~\cite{raffel2020exploring}), encoder-only (such as BERT~\cite{devlin2018bert}), and decoder-only (such as GPT~\cite{achiam2023gpt} and LLaMA~\cite{touvron2023llama}). Due to the advantages of higher training efficiency and better inference performance, state-of-the-art models predominantly use the decoder-only architecture. In this paper, we particularly focus on decoder-only LLMs.

\subsection{Tokens and Next Token Prediction}
\label{subsec:tokens}

\noindent
\textbf{Tokens:} 
Tokens serve as the fundamental building blocks that LLMs use to represent and interpret language. Tokenization is a critical step in preparing text for processing by the model. Before handling raw text input, the LLM employs a tokenizer to break the text or code into smaller units known as tokens. These tokens can represent anything from a single character to a whole word, depending on the complexity of the language and the tokenization method. For example, when an LLM processes code, it tokenizes the source code into units such as keywords, identifiers, operators, and literals. Each component of the code is treated as a token, which the model processes individually. 
Similarly, during the output process, the LLM generates tokens instead of producing full text all at once. The tokenizer then converts these tokens back into readable text or code.  The use of tokenization enables LLMs to handle different languages and contexts, regardless of their specific syntax, by converting text or code into tokens that can be systematically processed.

\vspace{2mm}
\noindent
\textbf{Next Token Prediction:} 
Next token prediction is a core task in training and operating LLM. It involves predicting the next token in a sequence of text based on the preceding tokens. In this process, the model takes an input sequence and generates the most likely next token, continuing this until the desired output length is reached. 
The training objective of the next token prediction can be formalized by the following loss function:
$$
L(\theta) = - \sum_{t=1}^{T} \log P(x_t | x_{<t}; \theta)
$$

Here, the goal is to minimize the negative log-likelihood to accurately predict the next token based on the given input. To generate an entire sentence, the model starts with a given initial token or prompt. It predicts the next token in the sequence using the learned probabilities, appends this token to the sequence, and then uses the extended sequence as input to predict the subsequent token. This process is repeated iteratively until a complete sentence is formed.

\subsection{Self-Attention Mechanism}
\label{subsec:attention}

The core of decoder layers is the self-attention mechanism, which allows a model to calculate the importance of different tokens in a sequence when encoding a specific token. When generating the next token, the LLM first calculates the relevance between the current token and all previous tokens. This relevance, known as the attention weight, is obtained through the self-attention mechanism. The model then uses these attention weights to perform a weighted sum of the values of all previous tokens, resulting in the value for the new token.

In decoder-only LLMs, to maintain the auto-regressive property during the generation process—where each token is generated based solely on the previously generated tokens without access to subsequent tokens—masks are applied during the calculation of attention weights. This ensures that the attention weights for the current token to any subsequent tokens are set to zero.
Furthermore, to enable parallel computation of self-attention, the multi-head attention mechanism divides each embedding into multiple parts, each processed by separate attention heads. The results from these heads are then concatenated and linearly transformed to produce the final output. Consequently, this approach allows the model to simultaneously attend to different aspects of the input sequence, thereby enhancing its ability to capture diverse features.

Attention captures the relevance between tokens, serving as an inherent explainability feature of the transformer architecture. There has been significant research~\cite{ghaeini2018interpreting, lee2017interactive, wiegreffe2019attention, serrano2019attention, wanWhatTheyCapture2022, ma2024unveiling} on leveraging attention for the interpretability of language models. 
Previous research~\cite{wanWhatTheyCapture2022, ma2024unveiling} has shown that the self-attention mechanism in LLMs can capture both syntactic and semantic information in code. Specifically, this is achieved by analyzing the attention distribution to identify syntactic or semantic relationships between tokens.

\section{Motivation, Key Idea, and Threat Model}

In this section, we first discuss the motivation driving our work (\S\ref{subsec:motivation}), then introduce our key idea, supported by a series of experiments (\S\ref{subsec:ee}). Lastly, we present our threat model (\S\ref{subsec:threatmodel}).

\subsection{Motivation}
\label{subsec:motivation}

In this section, we discuss the motivations of our work. As shown in \autoref{code:motivation}, we present a vulnerable program where, \textit{in line 3}, the function calculates the value of \texttt{len} based on the input \texttt{s} and subsequently allocates memory of size \texttt{len * sizeof(cairo\_glyph\_t)} in \textit{line 8}. However, the \texttt{gmalloc} function used for memory allocation is unsafe, as it lacks checks for integer overflow. If an overflow occurs when calculating \texttt{len * sizeof(cairo\_glyph\_t)}, it can lead to a buffer overflow, which could potentially be exploited for arbitrary code execution. The goal of vulnerability localization is to take code containing vulnerabilities as input and output the line in the code where the vulnerability resides, which in this example would be \textit{line 8.}

\begin{figure}[h]
\centering
\begin{minted}[
linenos,bgcolor=yellow!10,
xleftmargin=.05\textwidth,breaklines,
fontsize=\footnotesize,
highlightlines={8},
highlightcolor=pink
]{C}
void CairoOutputDev::beginString(GfxState *state, GooString *s)
{
    int len = s->getLength();
    if (needFontUpdate)
        updateFont(state);
    if (!currentFont)
        return;
    glyphs = (cairo_glyph_t *) gmalloc (len * sizeof (cairo_glyph_t));
    glyphCount = 0;
}
\end{minted}
\caption{Vulnerable code example}
\label{code:motivation}
\end{figure}

\begin{figure}[h]
\centering
\includegraphics[width=0.9\hsize]{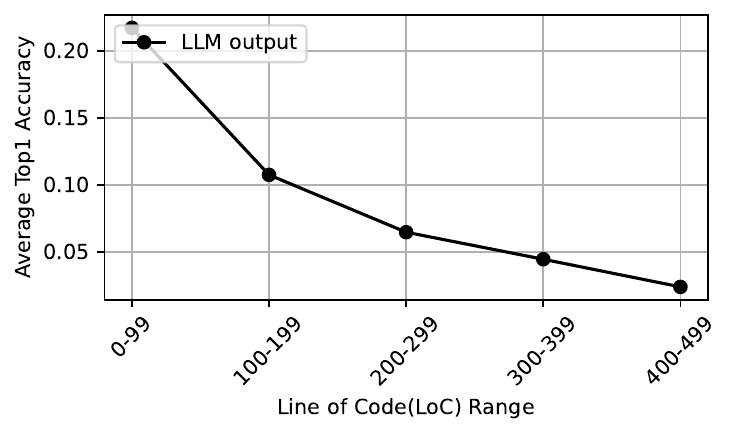}
\vspace*{-3mm}
\caption{The relationship between Lines of Code (LoC) and Average Top-1 Accuracy, generated by Llama-3.1-8B-Instruct, on all datasets in \S\ref{subsec:experimentsetup}}
\label{fig:motivation}
\end{figure}

\begin{figure}[h]
\centering

\subfloat[Non-vulnerable Program]{%
    \includegraphics[width=0.45\hsize]{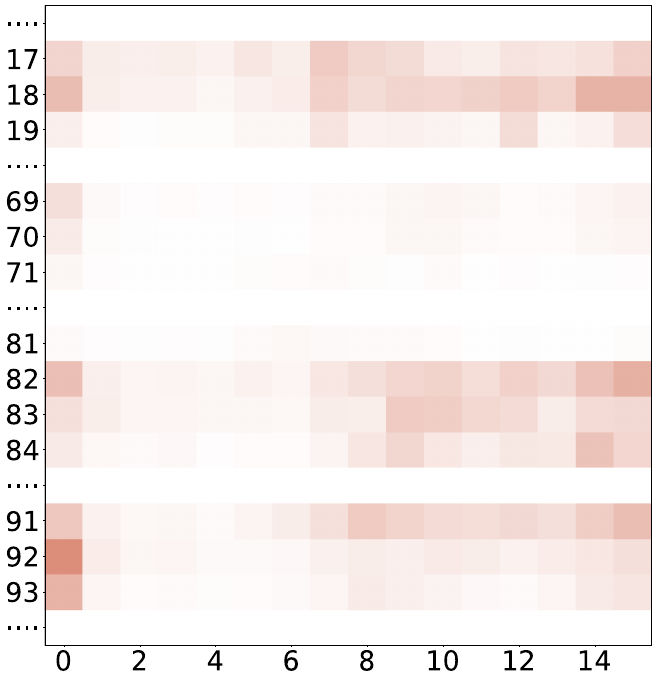}%
    \label{fig:motivation_attention_FF}
}\hfill
\subfloat[Highlighted Non-vulnerable Program]{%
    \includegraphics[width=0.45\hsize]{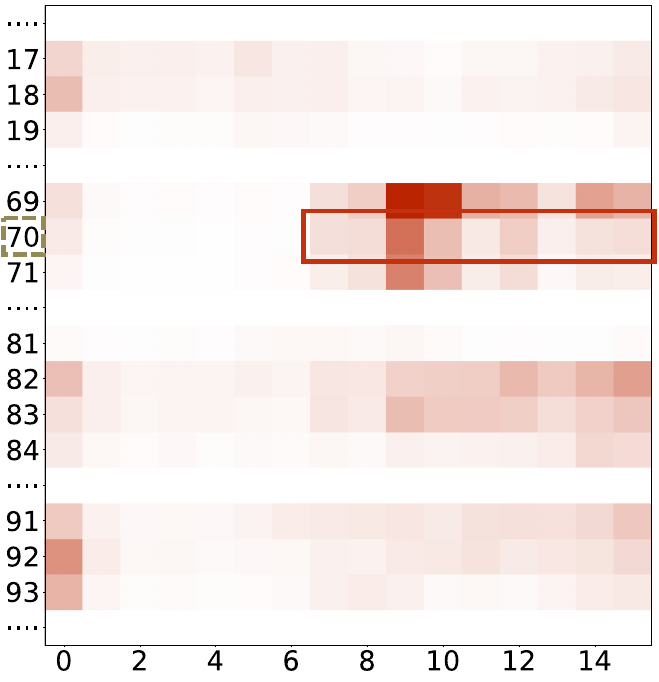}%
    \label{fig:motivation_attention_FT}
}
\vspace{0.2cm}

\subfloat[Vulnerable Program (injected at line 70)]{%
    \includegraphics[width=0.45\hsize]{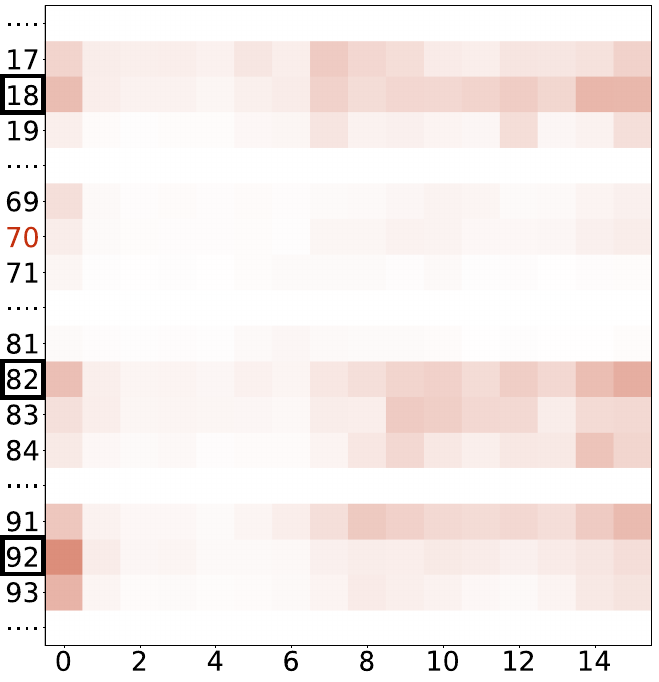}%
    \label{fig:motivation_attention_TF}
}\hfill
\subfloat[Highlighted Vulnerable Program]{%
    \includegraphics[width=0.45\hsize]{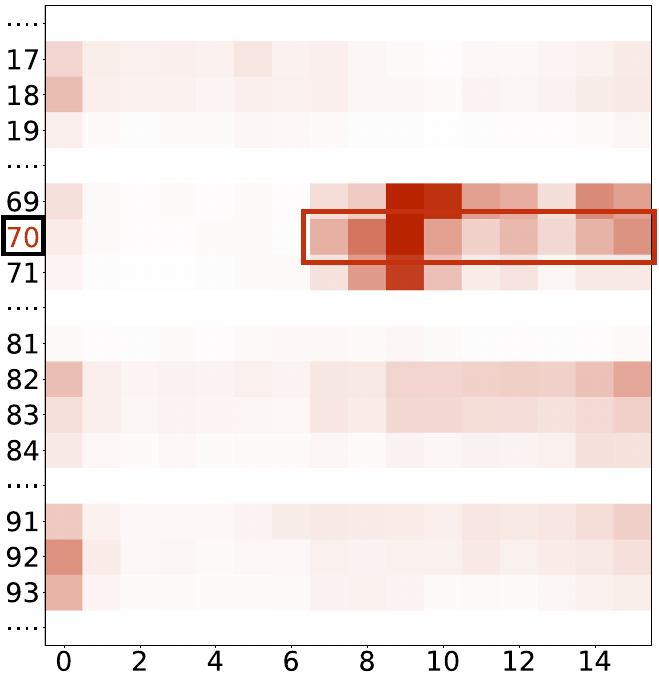}%
    \label{fig:motivation_attention_TT}
}

\caption{Attention maps for four programs, where (a) represents a vulnerability-free program, (c) shows the program after injecting a vulnerability at line 70 of (a). (b) and (d) represent the programs after highlighting line 70 for (a) and (c), respectively. In each figure, each row corresponds to a line number in the code, while the columns indicate the attention from different decoder layers.
Darker colors indicate stronger attention, and the black boxes on the line numbers mark the vulnerable lines identified by the LLM.}
\label{fig:motivation_attention}
\end{figure}

While it is true that directly querying LLMs to localize code vulnerabilities, or employing more advanced prompt engineering techniques such as Chain of Thought~\cite{wei2022chain, nong2024chain} (i.e., we can apply a step-by-step thought process to systematically break down the potential issues in code) can potentially produce the desired outcome, these methods are constrained by several significant limitations.

First, recent studies \cite{liu2024lost, kuratov2024babilong} have demonstrated that the reasoning capabilities of LLMs deteriorate substantially when working with longer contexts. Since vulnerability localization is inherently a reasoning task that often involves analyzing extended code segments, this degradation poses a significant challenge. For example, for the code shown in \autoref{code:motivation}, when we input the complete file containing the vulnerable code into the LLM for vulnerability localization, only two out of ten attempts successfully identified the vulnerability. Furthermore, as shown in \autoref{fig:motivation}, as the lines of code (LoC) increase, the output accuracy of the LLM declines significantly. When the LoC exceeds 300, the accuracy drops to below 5\%. While existing advanced reasoning methods (e.g., Self-Consistency~\cite{wang2022self} and rStar~\cite{qi2024mutual}) can improve reasoning to some extent, they are not fundamentally designed to address long context degradation. Consequently, there is still a significant decline in accuracy when dealing with long contexts.

Second, the advanced reasoning methods mentioned earlier typically employ stochastic decoding strategies to explore different reasoning paths, thereby enhancing accuracy. However, this approach introduces randomness into the output, leading to variability that can result in unstable localization outcomes. Consequently, it becomes challenging to pinpoint the exact line containing the vulnerability. 
Even when methods like majority voting or scoring are used to aggregate multiple outputs, the overall improvement in accuracy is often marginal~\cite{kang2024quantitative, purba2023software}.

\subsection{Exploratory Experiments and Key Idea}
\label{subsec:ee}

\noindent\textbf{Exploratory Experiments (EE) }  As discussed in \S\ref{sec:llm}, the core of LLMs' internal computation lies in the self-attention mechanism. By calculating attention weights between each pair of tokens, hidden states are derived, which are then used to generate the next token. Consequently, the output of LLMs is intrinsically tied to the attention weights, making self-attention one of the most widely used methods for interpreting LLM behavior. 
Therefore, \textit{we infer that understanding the relationship between attention weights and the LLM’s output could potentially help identify specific vulnerability locations (e.g., lines of code containing vulnerabilities may receive higher attention weights).}  

Specifically, we have formulated four hypotheses to enhance vulnerability localization in code. First, we hypothesize that accurate localization of a vulnerable line of code necessitates the model's focused attention on that specific line. Given the challenges posed by extended contexts, vulnerable lines may be overlooked. To address this, our second hypothesis proposes manually intensifying the model’s focus on identified vulnerable lines to improve detection accuracy. 
Thirdly, we posit that a genuinely vulnerable line will inherently exert a significant influence on the model's output. However, a major challenge is the inability to pre-identify which lines warrant this increased attention due to unknown vulnerability statuses. Intuitively, we suggest directing the model to scrutinize every line of code, irrespective of its vulnerability status. This approach hinges on our fourth hypothesis: intensifying the model’s focus across all code lines is unlikely to lead to incorrect evaluations, and focusing on non-vulnerable code will not adversely affect the attention dynamics. \looseness=-1

To validate our hypotheses, we conducted four exploratory experiments. To be more specific, we initially prepared a code snippet free of vulnerabilities, then intentionally introduced a vulnerability at a specific line. We tasked the LLM with identifying the vulnerable line of code. Notably, most large Transformer-based models (e.g., BERT, GPT) generate attention weights during training or inference, which indicate the model’s focus on different input positions across layers and attention heads. These attention weights can be extracted through APIs, enabling us to observe how attention shifts across layers and heads when processing different inputs. 
The experimental setup and results are discussed as follows: 

\vspace{2mm}
\noindent\textbf{EE-I:} We first queried the LLM without highlighting any specific lines.   The results are shown in \autoref{fig:motivation_attention} (a) and (c), where  \autoref{fig:motivation_attention} (a) illustrates the attention distribution for the code without vulnerability, and \autoref{fig:motivation_attention} (c) shows the attention for code with an injected vulnerability at line 70. Each row in the figures represents different lines of code,   while the columns correspond to various decoder layers. We observed that the LLM outputted lines 18, 82, and 92 as containing vulnerabilities, which is a mistaken judgment.   In the attention distribution displayed in \autoref{fig:motivation_attention} (c), these lines received more attention than others. In contrast, line 70, where the actual vulnerability exists, did not attract significant attention. This tendency for key content to be overlooked is particularly evident in longer scenarios.

\vspace{1.5mm}

 \begin{mdframed}[backgroundcolor=orange!4] 
 
\noindent\textit{\textbf{Observation (I).}{ The lines identified as vulnerable by the model are consistently highlighted in the attention distribution. However, insufficient attention may prevent the model from effectively detecting the vulnerability. As a result, the output may be prone to high false positives.}}
\end{mdframed}

\vspace{1.5mm}
\noindent\textbf{EE-II:}
We then explored how to direct the model’s attention toward the lines containing vulnerabilities. One straightforward approach is to modify the prompt to guide the model's focus to a specific line(i.e., to highlight this line). By instructing the model to ``\textit{pay attention to line 70}'', we successfully shifted its focus. The attention results, shown in \autoref{fig:motivation_attention} (b) and (d), indicate a significant increase in attention on line 70.

\vspace{1.5mm}
 \begin{mdframed}[backgroundcolor=orange!4] 
\noindent\textit{\textbf{Observation (II).} By adjusting the prompt, we can direct the model to focus on a specific line, effectively altering its attention.}
\end{mdframed}

\vspace{1.5mm}
\noindent\textbf{EE-III:}
We analyze the LLM's output (not the attention distribution) after applying highlighting. For the code without a vulnerability, even when highlighting line 70, the LLM did not mistakenly flag it as a vulnerability, suggesting that the highlighting did not cause hallucination issues. 
Conversely, for the code with a vulnerability, highlighting line 70 enabled the model to successfully identify the code as vulnerable.
 
\vspace{1.5mm}
 \begin{mdframed}[backgroundcolor=orange!4] 
\noindent\textit{\textbf{Observation (III).} By increasing attention to a vulnerable line, we can make it easier for the model to identify it as vulnerable.}
\end{mdframed}

\vspace{1.5mm}
\noindent\textbf{EE-IV:}
We then analyze the LLM's behavior when highlighting a non-vulnerable line, as shown in  ~\autoref{fig:motivation_attention} (b) and (d). The attention distribution for the highlighted line is shown within the red box, we observe that highlighting a vulnerable line leads to a significant increase in attention on that line. In contrast, when a non-vulnerable line is highlighted, the corresponding attention shows only a limited increase.

\vspace{1.5mm}
 \begin{mdframed}[backgroundcolor=orange!4] 
\noindent\textit{\textbf{Observation (IV).} When a vulnerable line is highlighted, certain attention patterns show significant increases, which do not appear for non-vulnerable lines.}
\end{mdframed}

\vspace{1.5mm}
 
\noindent\textbf{Key Idea:}  
As outlined in Observations (I)-(IV), LLMs tasked with vulnerability localization in extensive contexts frequently miss the actual vulnerable lines (O-I). By strategically modifying the model's prompts, we can enhance the model's focus on these vulnerable lines (O-II), thereby refining its capability to accurately localize vulnerabilities (O-III). Furthermore, directing the model's attention towards non-vulnerable lines has minimal impact on its overall attention allocation (O-IV). Therefore, our idea is that we can systematically iterate over each line of code, individually highlighting them to ensure comprehensive coverage of all potential vulnerabilities. This method is effective because highlighting a vulnerable line results in a marked increase in the model’s attention, whereas non-vulnerable lines exhibit minimal changes. By monitoring these attention shifts, we can precisely identify and confirm the presence of vulnerabilities, thus significantly improving the accuracy of vulnerability localization.

\subsection{Threat Model}
\label{subsec:threatmodel}

\noindent\textbf{Scope and Assumptions:} We now discuss the scope and assumptions of our work. The primary goal is to localize the vulnerabilities within a program, guided by the following requirements.  First, there are indeed numerous vulnerability localization tools in the software security domain, but our tool specifically focuses on those powered by LLMs, recognizing the immense potential they offer (e.g., CoT~\cite{wei2022chain}). Second, this paper focuses on the task of vulnerability localization within a single file. We do not address vulnerabilities that span across multiple files, such as those involving inter-file code dependencies or cross-file data flows. For example, vulnerabilities related to privacy leakage often occur when sensitive data is collected in one file, processed in another, and then improperly exposed or transmitted in yet another file. This is out of our focus. It is important to emphasize that our work focuses on vulnerability localization~\cite{xu2024racing, shen2021localizing, yin2024multitask} rather than vulnerability detection, meaning that all input programs are expected to contain at least one vulnerability. Our approach helps developers quickly pinpoint the location of vulnerabilities when a vulnerability trigger or crash is identified.

\vspace{1mm}
\noindent\textbf{Problem Formalization:}
We assume the vulnerable code is represented as \( C = [c_1, c_2, \ldots, c_n] \),  where \( c_i \) denotes the \( i \)-th line of code. In the context of vulnerability localization, we only consider inputs that contain vulnerabilities, meaning that the vulnerable code must include vulnerable lines \( V = [v_1, v_2, \ldots, v_k] \), where \( k \leq n \) and \( v_i \in C \).  The objective of vulnerability localization is to determine whether each line \( c_i \) is vulnerable, ultimately producing an output of the vulnerable lines \( O = [o_1, o_2, \ldots, o_m] \), where \( m \leq n \) and \( o_i \in C \).\looseness=-1

\section{Design of \tool}

\begin{figure*}
\centering
\includegraphics[width=\textwidth]{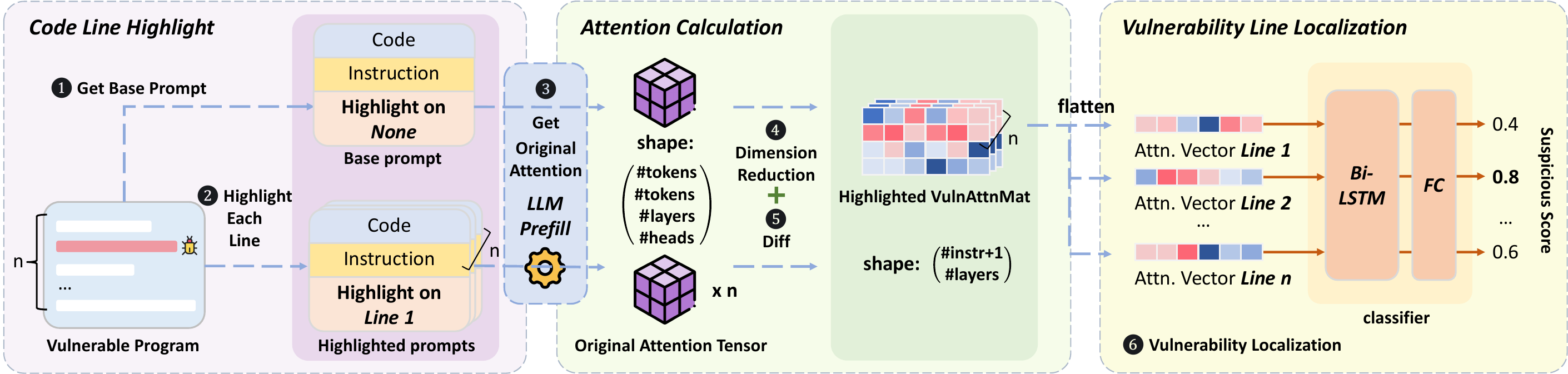}
\caption{Overview of \tool. For vulnerable code, the process begins by highlighting each line individually to generate a base prompt and a set of highlighted prompts. These prompts are then processed through the LLM's prefill phase to obtain their respective attention maps. For each highlighted attention map, the difference with the base attention map is computed, and the result is flattened to form an attention vector corresponding to each highlighted line. These attention vectors are then classified by a Bi-LSTM classifier to determine whether each line contains a vulnerability, ultimately yielding a suspicious score for each line.\looseness=-1}
\label{fig:overview}
\end{figure*}

In this section, we now discuss the design details of \tool. As shown in \autoref{fig:overview}, \tool consists of three primary stages, i.e., code line highlight, attention calculation, and vulnerability line localization. Specifically, in each of these steps, we first outline the challenges (\textit{{C-I}, {C-II}, {C-III}}) encountered, followed by the corresponding solutions (\textit{{S-I},  {S-II},  {S-III}}).
\begin{itemize}[partopsep=2pt, topsep=-\parskip, parsep=2pt, leftmargin=*]
    \item \textbf{Code Line Highlight (\S\ref{sub:codelinehighlight}):} In this stage, \tool generates two versions of the input based on the vulnerable program for LLM processing. The first is an unaltered copy with no line highlighted, while the second systematically highlights each line. 
    \item \textbf{Attention Calculation (\S\ref{sub:attentioncaculation}):} In this stage, \tool uses the LLM to generate attention maps for both the base and highlighted versions of the prompt. By comparing these maps, it assesses the impact of highlighting specific lines on the model's attention distribution. 
    \item \textbf{Vulnerability Line Localization (\S\ref{subsec:vulloc}):} In this stage, \tool analyzes the distribution of the model's attention to assess whether a specific line contains a vulnerability. By examining how strongly the model focuses on different parts of the code, \tool can effectively determine which lines are more likely to exhibit security issues. 
\end{itemize}

\subsection{Code Line Highlight}
\label{sub:codelinehighlight}

In this step, \tool generates two versions of the input based on the vulnerable program, both of which will be processed by the LLM in subsequent steps. The first version is a direct copy of the vulnerable program with the template prompt inserted, but none of the lines are highlighted (\textbf{Step \ding{182}}). The second version highlights each line of the vulnerable program (\textbf{Step \ding{183}}). To create the second version, \tool systematically examines each line of code, sequentially applying highlights. This highlighting is intended to direct the LLM’s focus to specific lines.
By highlighting a code line, the attention distribution across the prompt is altered (\textbf{Step \ding{184}}). \looseness=-1

\vspace{2mm}
\noindent\textbf{\textit{(C-I) Inappropriate prompt may mislead LLM:}} At first glance, highlighting a specific line may seem straightforward, with several methods available to achieve this. For instance, comments like \texttt{//Pay attention to this} or markers such as \texttt{/*FOCUS\_BEGIN*/} and \texttt{/*FOCUS\_END*/} can be inserted into the code to draw attention to a specific line~\cite{jiang2023impact, prenner2022can}. Another approach is to mask unimportant sections of the input, allowing the model to focus exclusively on the critical parts~\cite{berabi2024deepcode}. 

However, directly highlighting a line without considering domain-specific factors can introduce complications and potentially degrade the performance of \tool.  For example, inserting comments or markers in lengthy contexts may cause these markers to be overlooked due to the sheer amount of surrounding text, which could lead to poor attention enhancement. Imagine highlighting a line within a 200-line code file—comments such as \texttt{//Pay attention to this} might go unnoticed (as discussed in \S\ref{subsec:motivation}), particularly in larger contexts. Also,  while specifying areas of focus in the instructions can stabilize attention increases, it may lead to hallucinations in the language model. This can cause the model to mistakenly classify a correct line as vulnerable, especially when the model strongly focuses on that line, thereby failing to reduce the false positive rate effectively. Similarly, masking irrelevant sections can compromise the integrity of the code. For instance, masking entire functions deemed unimportant could prevent a valid comparison of attention differences across highlighted lines, hindering the model’s understanding of the overall logic.

\vspace{1mm}
 \begin{mdframed}[backgroundcolor=green!2] 
\noindent\textit{\textbf{(S-I) Highlight lines using their index.} 
Our approach is straightforward yet effective. Instead of adding extra content that might confuse the LLM, we directly highlight specific lines by specifying their line index. This method accurately enhances attention on the highlighted lines and improves the LLM's success rate in locating them. Additionally, the differences between lines are significant and easily analyzable, providing clear, quantifiable reference points for future analysis.
}
\end{mdframed}

\vspace{1mm}
Next, the key question is how to construct an effective prompt based on \textbf{(S-I)}. As depicted in \autoref{lst:prompt}, we have implemented the following steps: First, we prepend line indices to the code and instruct the model to "pay attention to" a specific index, thereby enhancing focus on that particular line. These indices are based on absolute positions, rather than relative ones, as LLMs have difficulty with relative positioning~\cite{zhang2024attention}. Second, our objective is to direct the model’s attention to the highlighted line and prompt it to determine whether that line contains a vulnerability, without introducing any prior assumptions. To this end, we instruct the LLM with the primary objective of identifying potential vulnerabilities in the entire code, while focusing specifically on the correctness of the highlighted line, rather than only judging the correctness of that line. Finally, to facilitate future analysis, we minimize the inclusion of non-code lines in the prompt. Since highlighting primarily impacts the attention directed toward the highlighted line and the instructions, reducing non-code lines helps streamline subsequent analyses, making it easier to track the model's attention shifts and performance.

\begin{figure}[H]
\definecolor{deepyellow}{rgb}{1.0, 0.9, 0.5}
\centering



\begin{subfigure}[H]{0.48\textwidth}
\begin{minted}[
bgcolor=yellow!10,
xleftmargin=.02\textwidth,
breaklines,
fontsize=\footnotesize,
highlightlines={3},
highlightcolor=deepyellow,
escapeinside=||
]{C}
Code: 
    ....
8:      glyphs = (cairo_glyph_t *) gmalloc (len * sizeof (cairo_glyph_t));
9:      glyphCount = 0;
10: }

Pay attention to line {|\textbf{8}|}. Check whether there are vulnerabilities in it.
vulnerable line: ```
\end{minted}
\vspace{-8mm}
\label{lst:file2}
\end{subfigure}

\caption{ 1 out of 10 highlighted prompt examples for one code (highlighting line 8)}
\label{lst:prompt}
\end{figure}

\subsection{Attention Calculation}
\label{sub:attentioncaculation}

After generating the base prompt and its corresponding highlighted version for each line, \tool utilizes the LLM to compute the attention maps for both sets of prompts. By calculating the differences between the attention maps of the highlighted and base prompts, \tool quantitatively assesses the influence of highlighting individual lines on the model's attention distribution.

\vspace{2mm}
\noindent\textbf{\textit{(C-II) Impracticality of analyzing large attention tensors:}} 
In the original attention output, a tensor with the shape \texttt{(num\_tokens,\ num\_tokens, num\_layers, num\_heads)} is generated, representing the attention scores calculated for every pair of tokens across all layers and heads of the model. This structure captures the intricate relationships between tokens and provides valuable insights into how attention is distributed throughout the network: 
\begin{itemize}
    \item \textbf{The first \texttt{num\_tokens}} represents the number of tokens, which is typically the number of the segment of code (or the instructions) after tokenization.  \looseness=-1
 \item \textbf{The second \texttt{num\_tokens}} represents the relationship of each token in the code (or text) with all other tokens that follow it (the first \texttt{num\_tokens}). \looseness=-1
 \item \textbf{\texttt{num\_layers}} refers to the different layers in the model. Each layer models the relationships between tokens more deeply, capturing different features of the code. \looseness=-1
 \item \textbf{\texttt{num\_heads}} refers to how many independent attention heads exist per layer. Each head focuses on different aspects of the code's structure and semantics. \looseness=-1
\end{itemize}

However, directly utilizing this tensor for assessing the impact of highlighting specific tokens or phrases proves to be highly impractical due to its sheer size. 
For example, with an input consisting of 1000 tokens, 32 decoder layers, and 32 attention heads, the resulting tensor would contain at least 0.5 billion elements.   Analyzing such a large tensor becomes computationally prohibitive, as it not only requires substantial memory but also complicates the extraction of meaningful insights regarding the effects of highlighting.

\vspace{1mm}
 \begin{mdframed}[backgroundcolor=green!2] 
\noindent\textit{\textbf{(S-II) Dimensionality reduction for vulnerability-focused attention analysis.} 
Our approach aims to reduce the dimensionality of the attention output, generating a tensor that not only captures the semantic information relevant to vulnerability localization but also contains fewer elements, thereby simplifying subsequent analysis.
}
\end{mdframed}

\vspace{1mm}

Please note that traditional dimensionality reduction methods are not suitable for this context, as they generally aim to preserve similarities from the high-dimensional space when reducing to lower dimensions. In contrast, our scenario focuses on extracting only the relevant components for comparison, rather than maintaining high-dimensional similarities. Additionally, the original attention outputs differ in size due to varying token counts, meaning there is no inherent similarity to retain in the high-dimensional space. Below, we outline the necessary steps for this stage.

\vspace{2mm}
\noindent\textbf{(Step \ding{185}) Dimensionality Reduction of Tensor:}
In this step, our method takes the original tensor with the dimensions \texttt{(num\_tokens, num\_tokens, num\_layers, num\_heads)} and systematically reduces it into a more manageable matrix, which we refer to as \textsf{LayerwiseAttnMat}. This matrix effectively captures essential attention patterns across layers and heads, streamlining further analysis.\looseness=-1

\begin{itemize}
    \item \textbf{Attention Head Aggregation (Step \ding{192}).} We combine the attention from all heads by summing them together. This merges the information captured by the different heads into one attention map with the shape \texttt{(num\_tokens, num\_tokens, num\_layers)}. Each attention head focuses on different features and relationships between tokens, so combining them helps gather a more complete understanding of the input.
    \item \textbf{Multi-Layer Attention Preservation (Step  \ding{193}).} The attention from all layers is retained because each decoder layer possesses a unique capacity to capture different semantic and relational features. For instance, lower layers may concentrate on fundamental patterns such as syntax, while higher layers tend to capture more abstract concepts, like the overall meaning of the code~\cite{ma2024unveiling}. By preserving attention across all layers, we ensure the model captures both basic and complex information, enriching the representation.
    \item \textbf{Last Token Attention for Context Aggregation (Step  \ding{194}).} The last token in the prompt is used as a query to determine how much attention it gives to all the preceding tokens, resulting in a matrix of size \texttt{(num\_tokens, num\_layers)}. This is important because, in decoder-only LLMs, attention is computed sequentially, meaning each token only attends to the tokens that came before it. By using the last token as the query, the model can aggregate all prior semantic information from the entire sequence. This approach helps the model make informed predictions or generate the next token, ensuring that the final token is informed by the full context of the prompt. 
    \item \textbf{Line-Level Attention Summation (Step \ding{195}).} The attention values of all tokens in each line of the prompt are added together to represent the overall attention for that line. This process results in a matrix of size \texttt{(num\_lines, num\_layers)}. We refer to this matrix as the \textsf{LayerwiseAttnMat}. The key idea here is that attention is inherently sparse—not all tokens receive equal focus~\cite{liu2022dynamic, liu2023deja}. Some tokens get significantly more attention than others. By summing the attention for each line, we simplify the data while still capturing the most important information, as many tokens contribute minimally to the overall attention distribution.
\end{itemize}

\vspace{2mm}
\noindent\textbf{(Step \ding{186}) Vulnerability-focused Matrix Generation:}
Next, \tool calculates the difference between the \textsf{LayerwiseAttnMat} of the highlighted prompt and the base prompt, resulting in the \textsf{DiffAttnMat}. This matrix represents how the model's attention changes when a specific line is highlighted. The purpose of \textsf{DiffAttnMat} is to quantify the effect of highlighting on the model's attention distributions. The difference in attention impact between vulnerable and non-vulnerable lines is most noticeable in two areas: the instruction section of the prompt (where the model receives guidance on how to process the input) and the highlighted line itself. Since these areas show the strongest influence, \tool focuses on the rows of the \textsf{DiffAttnMat} corresponding to the instruction section and the highlighted line, allowing it to accurately represent the attention shifts caused by highlighting. 
The final result is a matrix of size \texttt{([num\_lines\_of\_instr] + 1, num\_layers)}, known as the \textsf{VulnAttnMat}. This matrix captures the attention variations in a simplified and structured form, making it easier to analyze how the model distinguishes between vulnerable and non-vulnerable lines. The \textsf{VulnAttnMat} serves as the final output of this stage.

\begin{algorithm}
\setstretch{1}
 
\newcommand\Algphase[1]{%
\vspace*{-.5\baselineskip}\Statex\hspace*{\dimexpr-\algorithmicindent-2pt\relax}\rule{\hsize}{0.6pt}%
\Statex\hspace*{-\algorithmicindent}\textbf{#1}%
\vspace*{-.5\baselineskip}\Statex\hspace*{\dimexpr-\algorithmicindent-2pt\relax}\rule{\hsize}{0.6pt}%
}

\caption{Calculate \textsf{VulnAttnMat}}\label{vulnattnmat}
\begin{algorithmic}[1]
\Require Highlighted Prompt $P_{\text{highlight}}$, Base Prompt $P_{\text{base}}$, LLM $M_{\text{LLM}}$
\Ensure \textsf{VulnAttnMat} $V$ for the highlighted line

\Algphase{(Step \ding{185}) Dimensionality Reduction of Tensor}
\Function{getLayerwiseAttnMat}{$P, M_{\text{LLM}}$}
    \State $T \gets \text{tokenize}(P, M_{\text{LLM}})$
    \State $n \gets \text{countLines}(P)$
    \State $l, h \gets M_{\text{LLM}}.\text{num\_layers}, M_{\text{LLM}}.\text{num\_heads}$
    \State $M_{\text{tokens}} \gets \text{getOriginalAttention}(M_{\text{LLM}}, T)$ 
    \Statex \hspace{1.6em}{\textbf{Step \ding{192}:}}
    \State $M_{\text{tokens}} \gets \text{sum}(M_{\text{tokens}}, \text{axis}=3)$ 
    \Statex \hspace{1.6em}{\textbf{Step \ding{193} and \ding{194}:}}
    \State $M_{\text{last}} \gets M_{\text{tokens}}[-1][:][:]$ 
    
    \Statex \hspace{1.6em}{\textbf{Step \ding{195}:}}
    \State $\text{tokens\_for\_each\_line} \gets \text{splitTokensByLines}(T)$ 
    \State Initialize $A$ as an empty matrix of size $(n \times l)$
    
    \For{$line \in \{1, \dots, n\}$}
        \State $A[line] \gets \sum_{i \in \text{tokens\_for\_each\_line}[line]} M_{\text{last}}[i][:]$
    \EndFor

    \State \Return $A$
\EndFunction

\Algphase{(Step \ding{186}) Vulnerability-focused Matrix Generation:}
\State $\text{instruction\_indices} \gets \text{getInstrIndices}(P_{\text{highlight}})$
\State $\text{highlighted\_index} \gets \text{getHighlightedIndex}(P_{\text{highlight}})$
\State $A_{\text{highlighted}} \gets \text{getLayerwiseAttnMat}(P_{\text{highlight}}, M_{\text{LLM}})$
\State $A_{\text{base}} \gets \text{getLayerwiseAttnMat}(P_{\text{base}}, M_{\text{LLM}})$
\State \begin{varwidth}[t]{\linewidth}
      $V_{\text{highlighted}} \gets A_{\text{highlighted}}[\text{instruction\_indices} + \text{highlighted\_index}, :]$
      \end{varwidth}
\State $V_{\text{base}} \gets A_{\text{base}}[\text{instruction\_indices} + \text{highlighted\_index}, :]$
\State $V \gets V_{\text{highlighted}} - V_{\text{base}}$

\State \Return $V$
\end{algorithmic}
\end{algorithm}

\subsection{Vulnerability Line Localization}
\label{subsec:vulloc}

In this step, after obtaining the \textsf{VulnAttnMat} for each line of the vulnerable program, \tool determines whether the corresponding line contains a vulnerability or not through analyzing the attention patterns (\textbf{Step \ding{187}}).\looseness=-1

\noindent\textbf{\textit{(C-III) Precise Multi-Language Vulnerability Localization:}} We found that attention patterns for code or text within a single programming language (e.g., Java, C) often exhibit similarities. These patterns tend to reflect semantic information related to vulnerabilities at specific lines and layers in the code.  Therefore,  summing the attentions at these positions on the \textsf{VulnAttnMat} can yield a corresponding suspicion score (which is a numerical value assigned to a data point that quantifies how ``suspicious'' or anomalous it is) that highlights which sections of the code are more likely to contain vulnerabilities. However, vulnerabilities differ greatly across languages. For instance, in network protocols written in C, a buffer overflow might occur during packet processing, whereas in a web application, the vulnerability might involve cross-site scripting in JavaScript.  These varying contexts mean that attention patterns that indicate a vulnerability in one language may not be applicable to other languages. Additionally, these summed attention values serve as a coarse heuristic, providing a suspicion score that only ranks the relative suspiciousness of code lines, but falls short of achieving accurate vulnerability localization. While this can prioritize certain areas for review, it does not provide a probability for each line being vulnerable. For effective vulnerability localization, it's crucial to assess the actual probability of each line containing a vulnerability, rather than just ranking the lines by suspicion. \looseness=-1

\vspace{1mm}
 \begin{mdframed}[backgroundcolor=green!2] 
\noindent\textit{\textbf{ (S-III) Language-aware deep learning-based vulnerability localization.} 
We chose to employ a deep learning approach to classify the attention matrix and determine whether the lines contain vulnerabilities.  This  method can leverage complex patterns and relationships within the data to provide a more accurate and probabilistic assessment. Unlike simply ranking lines based on a suspicion score obtained by crudely summing attention values, a deep learning model can learn to identify specific vulnerability signatures in different languages, improving precision in vulnerability localization.
}
\end{mdframed}

\vspace{1mm}

The key question now is which deep learning method is most suitable for our context. In designing the learning algorithm, it was essential to account for the contextual dependencies within the classification task. Whether an attention matrix is deemed suspicious is influenced by the relationships among the attention matrices across all lines of code.

To address this, \tool implements a learning approach that integrates sequential information by employing a Bi-LSTM model, which effectively captures these contextual dependencies: First, The identification of anomalies of \textsf{VulnAttnMat} is context-dependent. For example, when inputting three lines of code, we generate three corresponding \textsf{VulnAttnMat}: $A$, $B$, and $C$. If $A$ significantly deviates from $B$ and $C$, it suggests that $A$ may be the most suspicious. To effectively model this, a sequence-to-sequence approach is necessary, with LSTM~\cite{sherstinsky2020fundamentals} and Transformer being the most common models for such tasks. Second, given that the features of anomalies in our case are relatively distinct, a straightforward LSTM model is sufficient. However, we opted for a Bi-LSTM~\cite{graves2013hybrid} to leverage its ability to capture bidirectional dependencies, as standard LSTM only processes information in one direction. This is important because the detection of anomalies is influenced by both the preceding and succeeding contexts, rather than solely relying on the prior context.

The training process begins by flattening the \textsf{VulnAttnMat} matrices corresponding to each vulnerable program into vectors, generating \texttt{num\_lines} vectors. These vectors are then fed into a Bi-LSTM network, followed by a fully connected layer, to perform binary classification at each time step. The model is optimized by minimizing the Binary Cross Entropy Loss, formulated as: 
    \[
    \mathcal{L} = -\frac{1}{N} \sum_{i=1}^{N} \left[ y_i \log(\hat{y}_i) + (1 - y_i) \log(1 - \hat{y}_i) \right]
    \]
    where \( N \) is the number of lines, \( y_i \) is the true label of the \( i \)-th line, and \( \hat{y}_i \) is the predicted probability of the line containing a vulnerability.
During the evaluation phase, the flattened vectors of the \textsf{VulnAttnMat} are input into the model to compute suspicious scores for each line. These scores are subsequently ranked in descending order to inform the vulnerability localization process, with scores greater than 0.5 being considered vulnerable lines.

\section{Evaluation}

In this section, we first outline the experimental setup (\S\ref{subsec:experimentsetup}), followed by an evaluation of \tool by addressing the following four research questions (\S\ref{subsec:result}).  

\vspace{2mm}
\begin{itemize}[left=1cm]
    \item [\textbf{RQ1.}] How accurately does \tool localize vulnerabilities compared to existing techniques (Effectiveness)? 
    \item [\textbf{RQ2.}] How effective is \tool when applied to different languages (Scalability)? 
    \item [\textbf{RQ3.}] How effective is \tool when applied to different LLMs (Robustness)? 
    \item  [\textbf{RQ4.}] How does the design of \tool impact the results (Ablation Study)?
\end{itemize}

\subsection{Experiment Setup}
\label{subsec:experimentsetup}

\noindent\textbf{Dataset:} Our dataset, comprises vulnerability benchmarks from C/C++ projects (\bigvul), Solidity smart contracts (\smartfix), and multi-language CVE records (\cvefixes). This allows for a comprehensive evaluation of \tool's performance across diverse contexts. Specifically:

\begin{itemize}
\item \textbf{\bigvul\cite{fan2020ac}}: A C/C++ code vulnerability benchmark collected from 384 open-source GitHub projects. \bigvul includes both vulnerable code snippets and their corresponding fixing commits. Vulnerabilities are annotated at function context level, categorizing the types of vulnerabilities and associated commits. \bigvul contains a total of 11,823 vulnerable functions.
\item \textbf{\smartfix\cite{so2023smartfix}}: A Solidity code vulnerability benchmark collected from multiple sources. \smartfix covers five common types of smart contract vulnerabilities: Integer Overflow, Reentrancy, Ether Leak, Suicidal, and incorrect use of `tx.origin`. Each vulnerability is annotated with its corresponding line numbers. \smartfix contains a total of 361 contract files.
\item \textbf{\cvefixes\cite{bhandari2021cvefixes}}: A multi-language code vulnerability benchmark that is automatically collected and curated from CVE records in the public U.S. National Vulnerability Database (NVD). \cvefixes collects both vulnerabilities and the commits that fix them, including 5,365 CVEs and 50,322 methods. \cvefixes encompasses multiple languages, such as C, C++, PHP, Python, Java, and JavaScript. From these, we selected the three most commonly used languages, C, Python, and Java, as benchmarks.
\end{itemize}

\bigvul provides vulnerability data at the function context level, \cvefixes provides vulnerability data at both the function and file context levels, and \smartfix provides vulnerability data at the contract context level. To evaluate \tool's performance across different contexts, we conduct evaluations at function, file, and contract context levels. Additionally, we filter out contexts exceeding 4,000 tokens to comply with the input constraints of LLMs.
In total, seven datasets were created, as shown in \autoref{table:datasets}. Among them, the C datasets \bigvul and \cvefixesc each contain 1,000 instances. For Java, there are two datasets: \cvefixesjm and \cvefixesj, corresponding to method-level and file-level contexts, respectively. In Python, the \cvefixespm and \cvefixesp datasets represent method-level and file-level contexts, respectively. Additionally, Solidity has one dataset, \smartfix, focusing on contract-level context.

\begin{table}[htbp]
    \centering
    \scriptsize
\setlength\tabcolsep{1.6pt}
   \caption{All datasets, C, J, and P, represent programs written in C, Java, and Python, respectively.}
    \begin{tabular}{lccccccc}
        \toprule[1pt]
        \multirow{2}{*}{\textbf{Context-level}} & \multicolumn{3}{c}{\textbf{Method}} & \textbf{Contract} & \multicolumn{3}{c}{\textbf{File (\cvefixes)}} \\
        \cmidrule(lr){2-4} \cmidrule(lr){6-8}
        & \bigvul & \cvefixesjm & \cvefixespm & \smartfix & \textit{C} & \textit{J} & \textit{P} \\
        \midrule
        \textbf{Count} & 1,000 & 500 & 500 & 349 & 1,000 & 500 & 500 \\
        \textbf{Avg LoC} & 60 & 23 & 29 & 119 & 230 & 168 & 203 \\
        \textbf{Avg Vuln lines} & 4 & 3 & 4 & 1 & 5 & 7 & 6 \\
        \bottomrule[1pt]
    \end{tabular}

    \label{table:datasets}
\end{table}

\vspace{2mm}
\noindent\textbf{Baseline:}
We primarily compare our approach with the direct outputs of LLMs. 
Following previous works \cite{zhang2024empirical, kangQuantitativeQualitativeEvaluation2024}, we provide the vulnerable code to an instruction-based LLM, which outputs all lines containing vulnerabilities and repeats the process \(k\) times. After \(k\) iterations, the suspicious score for each line is computed. The suspicious score for line \(m\) is calculated using the following formula:
\[
\operatorname{score}(m) = \frac{1}{k} \sum_{i=1}^{k} \left( \frac{1}{\left|r_{i}\right|} \cdot \left[ m \in r_{i} \right] \right)
\]
where \(r_i\) represents all the vulnerable lines output by the LLM in the \(i\)-th iteration.
In the case of ties, the lines are ranked according to the order in which they were output.
We set \(k=10\), meaning each piece of vulnerable code is processed by the LLM to output ten times.

\vspace{2mm}
\noindent\textbf{Metrics:}
 We use the following metrics: 
Top-N measures how many relevant items (e.g., correct localization) appear within the top N results returned by our system. Following previous works, we use $N=1, 3, 5$ to evaluate our approach~\cite{kangQuantitativeQualitativeEvaluation2024}. 
Precision measures the proportion of correctly identified positive instances out of all instances that were predicted as positive.  Recall, also known as sensitivity or true positive rate, measures the proportion of correctly identified positive instances out of all actual positive instances. F1-Score, which is the harmonic mean of Precision and Recall, provides a balanced metric that considers both false positives and false negatives. The F1-Score provides a single metric that balances the trade-off between precision and recall. The detailed definitions and formulations are omitted here, as they are extensively covered in other works.

\vspace{2mm}
\noindent\textbf{Model:}
All experiments, except for RQ3 (Robustness), were conducted on the Llama-3.1-8B-Instruct~\cite{touvron2023llama} model, which is one of the state-of-the-art LLMs with fewer than 10b parameters.

\vspace{2mm}
\noindent\textbf{Environment:}
The experiments were conducted on a computing system running Ubuntu 22.04.4 LTS. The system is equipped with an NVIDIA A800 80GB PCIe GPU with CUDA version 12.4 and an AMD EPYC 7763 processor.
\subsection{Experiment Results}
\label{subsec:result}

\vspace{2mm}
\noindent\textbf{Effectiveness (RQ1):}  
The goal of the experiment is to evaluate how accurately \tool localizes vulnerabilities compared to existing techniques. The evaluation is conducted using four different decoding-based settings of an LLM: (i) vanilla output, (ii) few-shot chain-of-thought (CoT) output~\cite{wei2022chain}, (iii) Mixture-of-Agents (MoA)~\cite{wang2024mixture} output and (iv) \textit{rStar}~\cite{qi2024mutual} output. In the vanilla output setting, the model directly generates the final result. In the few-shot CoT output, the model utilizes the few-shot CoT technique by first presenting reasoning examples to the LLM. The model then generates a step-by-step reasoning process before providing the final answer. This method enhances the model's reasoning capabilities. The MoA output requires the LLM to generate multiple results and then evaluate each output by themselves. Based on these evaluations, the model will provide the final output. Using MoA can help reduce the bias introduced by the randomness of the LLM. \textit{rStar} combines the reasoning and validation processes by first having the LLM generate multiple human-like reasoning actions to form different reasoning trajectories. Next, the LLM evaluates these trajectories to obtain the most mutually consistent trajectory as the final result. \textit{rStar} is currently one of the state-of-the-art methods for enhancing the reasoning capabilities of LLMs based on reasoning chain construction.  For each dataset, we performed 5-fold cross-validation to obtain results across the entire dataset.

\begin{table}[ht]
\centering
\scriptsize
\setlength\tabcolsep{4pt}
\caption{Comparison of Effectiveness Across Methods on C Language Datasets: \bigvul and \cvefixesc. \smartfix is excluded in this RQ as it is specific to Solidity language.}
\begin{tabular}{lrrrrrr}
\toprule[1pt]
\multirow{2}{*}{\textbf{Method}} & \multicolumn{3}{c}{\textbf{\bigvul}} & \multicolumn{3}{c}{\textbf{\cvefixesc}} \\
\cmidrule(lr){2-4} \cmidrule(lr){5-7}
& \textbf{Precision} & \textbf{Recall} & \textbf{F1} & \textbf{Precision} & \textbf{Recall} & \textbf{F1} \\
\midrule
\textbf{\tool}  & \textbf{37.6} & 13.2 & \textbf{19.5} & \textbf{29.9} & 10.5 & \textbf{15.5} \\
\midrule
Vanilla output & 7.8  & \textbf{43.4} & 13.1 & 2.4  & \textbf{45.1} & 4.6  \\
CoT output     & 20.4 & 5.8  & 9.0  & 7.4  & 1.8  & 2.9  \\
MoA output     & 8.5 & 38.2 & 13.9 & 2.9  & 32.2 & 5.4  \\
rStar output   & 10.8 & 18.5 & 13.7 & 4.1 & 11.0  & 6.0  \\
\bottomrule[1pt]
\end{tabular}
\label{table:main}
\end{table}

As shown in \autoref{table:main}, \tool achieved the highest F1-score on both the \bigvul and \cvefixesc datasets compared to the baseline. Specifically, on \bigvul, it improved by 1.5$\times$ and 2.2$\times$ over vanilla output and CoT output, respectively. On \cvefixesc, it achieved gains of 3.4$\times$ and 5.3$\times$ over vanilla output and CoT output, respectively. While vanilla output attained the highest recall, it had the lowest precision, indicating a higher likelihood of producing false positives. By examining the model's output, it can be seen that vanilla output often marks a large number of suspicious lines, leading to more false positives. In contrast, CoT output, after explaining the vulnerability, more precisely identifies the vulnerable lines, resulting in higher precision but lower recall. Although more advanced reasoning techniques like MoA and \textit{rStar} show improvement over CoT and Vanilla, they still perform significantly below \tool. \tool achieved the highest precision, striking a balance between precision and recall compared to the baselines, which led to the highest F1-score.

\begin{figure}[]
\centering
\subfloat[Comparison on \bigvul dataset]{
    \hspace{-3mm}
    \begin{minipage}[b]{\hsize}
    \includegraphics[width=\textwidth]{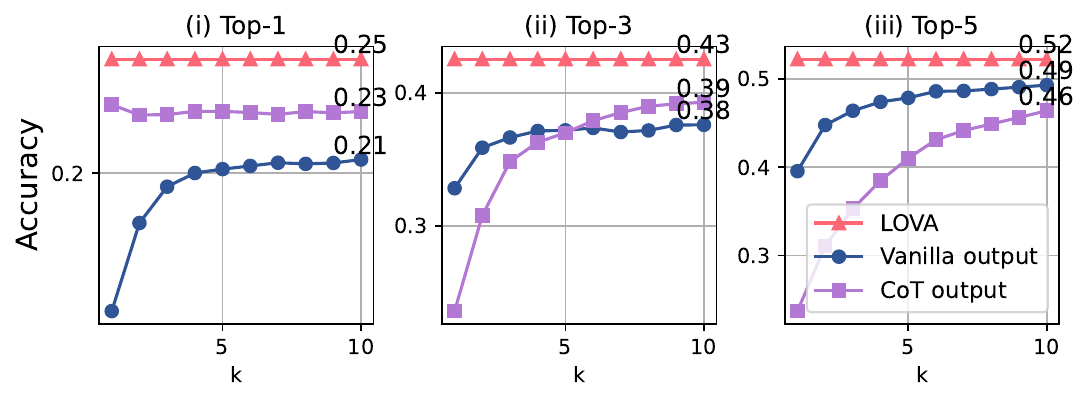}
    \end{minipage}
    }

\subfloat[Comparison on \cvefixesc dataset]{
    \hspace{-3mm}
    \begin{minipage}[b]{\hsize}
    \includegraphics[width=\textwidth]{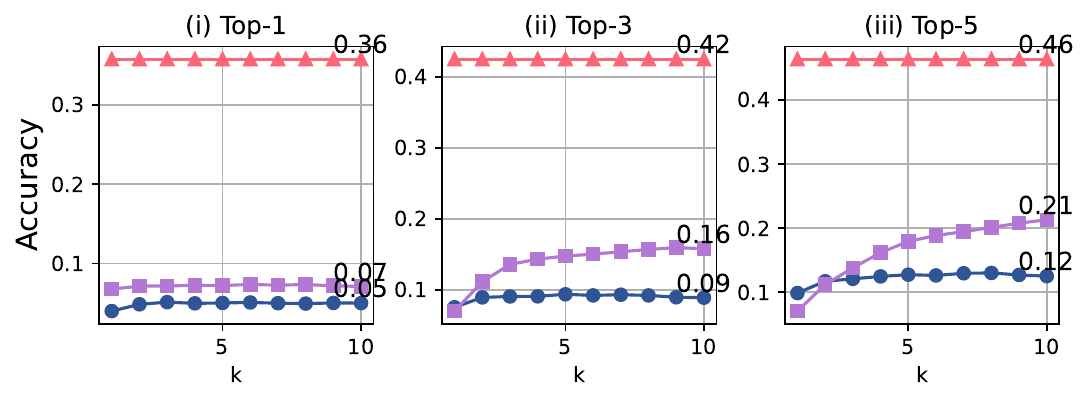}
    \end{minipage}
    }
\caption{Comparison of Top-N between \tool and baselines, where k refers to the iterations of the baseline.}
\label{fig:topn}
\end{figure}

\begin{figure}[h]
\centering
\includegraphics[width=0.9\hsize]{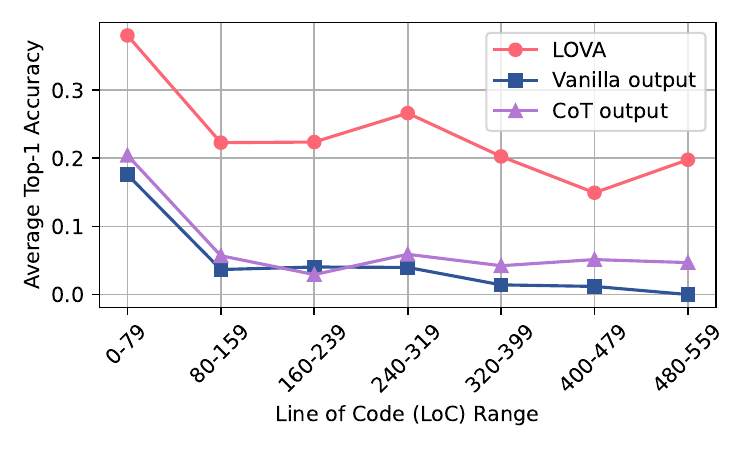}
\vspace*{-3mm}
\caption{Top-1 Accuracy on different line ranges}
\label{fig:line}
\end{figure}

In addition to F1-score, Top-N accuracy is also a useful metric for evaluating vulnerability localization. We compare the Top-N accuracy of \tool with both vanilla output and CoT output and investigate how combining multiple rounds of LLM outputs affects the accuracy. The experimental results are shown in \autoref{fig:topn}. The x-axis represents the accuracy after combining the results of k outputs, and the y-axis represents the Top-N accuracy, where N = 1, 3, 5. 
\tool demonstrates a marked improvement over both the vanilla and CoT outputs. In the Top-1, Top-3, and Top-5 categories, the vanilla output fails to generate accurate vulnerability localization results consistently. To further evaluate \tool’s performance in managing long contexts, we compared its effectiveness against vanilla and CoT outputs across different ranges of lines of code (LoC) using the \bigvul and \cvefixesc datasets, as shown in \autoref{fig:line}. The results indicate a substantial decline in the accuracy of both vanilla and CoT outputs as the LoC increases, with average Top-1 accuracy dropping below 10\% when the LoC exceeds 80 lines. In contrast, \tool consistently maintains high accuracy across varying LoC ranges. \looseness=-1

\vspace{2mm}
 \begin{mdframed}[backgroundcolor=red!2] 
 
\noindent\textit{\textbf{Answer (RQ1).} {\tool demonstrates exceptional vulnerability localization capabilities compared to the direct outputs of LLMs. In longer contexts, \tool performs significantly better, achieving up to a 5$\times$ improvement over the baseline on F1-score.}}
\end{mdframed}

\begin{table*}[ht]
\centering
    \scriptsize
\setlength\tabcolsep{10pt}
\caption{Comparison of F1 and Top-1 Scores Among Java, Python, and Solidity Languages}
\begin{tabular}{lrrrrrrrrrr}
    \toprule[1pt]
    & \multicolumn{4}{c}{\textbf{Java}} & \multicolumn{4}{c}{\textbf{Python}} & \multicolumn{2}{c}{\textbf{Solidity}} \\
    \cmidrule(lr){2-5} \cmidrule(lr){6-9} \cmidrule(lr){10-11}
    & \multicolumn{2}{c}{\textbf{\cvefixesj}} & \multicolumn{2}{c}{\textbf{\cvefixesjm}} & \multicolumn{2}{c}{\textbf{\cvefixesp}} & \multicolumn{2}{c}{\textbf{\cvefixespm}} & \multicolumn{2}{c}{\textbf{\smartfix}} \\
    \midrule
    & F1 & Top-1 & F1 & Top-1 & F1 & Top-1 & F1 & Top-1 & F1 & Top-1 \\
    \midrule
    \textbf{\tool}          & \textbf{21.3} & \textbf{181 (36.2\%)} & \textbf{34.6} & \textbf{281 (56.2\%)} & \textbf{18.1} & \textbf{98 (19.6\%)} & \textbf{31.0} & \textbf{204 (40.8\%)} & \textbf{80.0} & \textbf{284 (81.3\%)} \\
    Vanilla output & 9.0 & 54 (10.8\%) & 22.1 & 173 (34.6\%) & 6.5 & 66 (13.2\%) & 21.4 & 177 (35.4\%) & 5.5 & 90 (25.8\%) \\
    CoT output     & 3.2 & 59 (11.8\%) & 22.1 & 177 (35.4\%) & 4.7 & 77 (15.5\%) & 17.9 & 150 (30.0\%) & 35.1 & 134 (38.3\%) \\
    \bottomrule[1pt]
\end{tabular}

\label{table:multilang}
\end{table*}
\vspace{2mm}
\noindent\textbf{Scalability (RQ2):}  
One key advantage of LLMs over traditional smaller machine learning models lies in their ability to generalize across multiple languages. To evaluate the scalability, we conduct vulnerability localization in three additional languages—Python, Java, and Solidity—beyond C/C++. The results are shown in \autoref{table:multilang}. It is clear that \tool achieved the highest Top-1 accuracy and F1-score across all datasets. On the method-level-context datasets \cvefixesjm and \cvefixespm, \tool improved the F1-score by up to 1.6$\times$ and 1.7$\times$ compared to the baselines, respectively. In longer contexts, \tool achieved an impressive 14.6$\times$ improvement on the contract-level context dataset \smartfix compared to the vanilla output. The CoT output also achieved a high F1 score of 35.1 on this dataset. This may be due to the \smartfix dataset containing only a few simple types of vulnerabilities, making it easier for CoT to infer vulnerabilities. For the file-level-context datasets \cvefixesj and \cvefixesp, it achieved increases of up to 6.7$\times$ and 3.9$\times$, respectively, demonstrating \tool's remarkable ability to handle long contexts effectively. \looseness=-1

\vspace{1mm}

 \begin{mdframed}[backgroundcolor=red!2] 
 
\noindent\textit{\textbf{Answer (RQ2).} {\tool achieves excellent results in vulnerability localization across multiple programming languages, showing up to a 14.6$\times$ improvement over the baseline. This highlights \tool's impressive generalization capabilities.}}
\end{mdframed}

\vspace{2mm}
\noindent\textbf{Robustness (RQ3):}  
The robustness of \tool across various LLM is a critical factor in assessing its general applicability. To evaluate its effectiveness, we applied \tool to multiple state-of-the-art LLMs, including Llama-3.1-8B-Instruct~\cite{touvron2023llama}, Mistral-7B-Instruct-v0.2~\cite{jiang2023mistral}, and Phi-3.5-mini-instruct~\cite{abdin2024phi}, and measured its performance across vulnerability localization tasks. The results, summarized in \autoref{table:rq3}, show that \tool consistently outperforms baseline methods regardless of the underlying LLM. 
Specifically, when applied to Mistral, \tool improved the Top-1 accuracy by 1.1$\times$ compared to CoT outputs and achieved an F1 score that was 1.5$\times$ higher than vanilla outputs on method-level vulnerability localization tasks. On Phi, \tool exhibited relatively low accuracy across file-level vulnerability datasets. This may be due to Phi having only 3.8b parameters, the fewest among the three models. However, \tool still achieved a 3.4$\times$ increase in F1 score compared to CoT outputs, demonstrating its adaptability across models with varying parameter sizes. 
Notably, \tool’s performance on Llama was particularly impressive in handling longer contexts, where it outperformed CoT outputs by up to 4.9$\times$ in terms of Top-1 accuracy and showed a 3.4$\times$ increase in F1 score compared to the vanilla output on the file-level datasets. This demonstrates \tool’s robust capability to adapt to different LLM architectures and maintain high accuracy in diverse vulnerability localization scenarios.\looseness=-1

\begin{table}[ht]
\centering
\scriptsize
\setlength\tabcolsep{4pt}
\caption{Comparison of Methods Across C Datasets: \bigvul and \cvefixesc. \smartfix is excluded in this RQ as it is specific to Solidity language.}
\begin{tabular}{l l   r r   r r   r r}
\toprule[1pt]
\textbf{Datasets} & \textbf{Methods} & \multicolumn{2}{c}{\textbf{Llama}} & \multicolumn{2}{c}{\textbf{Mistral}} & \multicolumn{2}{c}{\textbf{Phi}} \\ 
\cmidrule(lr){3-4} \cmidrule(lr){5-6} \cmidrule(lr){7-8} 
& & \textbf{F1} & \textbf{Top-1} & \textbf{F1} & \textbf{Top-1} & \textbf{F1} & \textbf{Top-1} \\ 
\midrule
\multirow{3}{*}{\bigvul} & \textbf{\tool}   & \textbf{19.5} & \textbf{24.5} & \textbf{18.0} & \textbf{22.9} & \textbf{17.0} & \textbf{23.9} \\ 
                           & Vanilla output & 13.1  & 20.6  & 12.4  & 22.4  & 12.9  & 23.4  \\ 
                           & CoT output    & 9.0  & 23.0  & 10.5  & 21.2  & 8.2  & 14.7  \\ 
\midrule
\multirow{3}{*}{\cvefixesc} & \textbf{\tool}   & \textbf{15.5} & \textbf{35.8} & \textbf{14.0} & \textbf{32.7} & \textbf{11.4} & \textbf{29.0} \\ 
                           & Vanilla output & 4.6  & 5.1  & 4.7  & 6.5  & 4.9  & 4.8  \\ 
                           & CoT output    & 2.9  & 7.3  & 3.5  & 4.9  & 3.4  & 3.7  \\ 
\bottomrule[1pt]
\end{tabular}

\label{table:rq3}
\end{table}

 \vspace{2mm}
 \begin{mdframed}[backgroundcolor=red!2] 
\noindent\textit{\textbf{Answer (RQ3).} {\tool can be applied across various LLMs, as the self-attention mechanisms in different models consistently demonstrate semantic connections with the output.}}
\end{mdframed}
 \vspace{2mm}
\noindent\textbf{Ablation Study (RQ4):} 
To better assess the contribution of each design element, we conducted an ablation study focusing on three key components of the method: Code Line Highlight, Attention Calculation, and Vulnerability Line Localization. From this, we derived three distinct variants of \tool: \toolc, \toola, and \toolv. In \toolc, the code line highlighting mechanism was altered by introducing marks at specific positions that require emphasis. \toola utilizes an average pooling technique for dimensionality reduction and feature extraction, compressing the attention outputs into vectors of uniform length. Lastly, \toolv replaces the Bi-LSTM in vulnerability localization with a multilayer perception (MLP) network, omitting the use of context during the process.\looseness=-1

\vspace{-1mm}
\begin{table}[ht]
\centering
\scriptsize
\setlength\tabcolsep{5pt}
\caption{Accuracy comparison across \tool variants.}
\begin{tabular}{lrrrrrrrr}
\toprule[1pt]
\multirow{2}{*}{\textbf{Method}} & \multicolumn{4}{c}{\textbf{\bigvul}} & \multicolumn{4}{c}{\textbf{\cvefixesc}} \\
\cmidrule(lr){2-5} \cmidrule(lr){6-9}
& \textbf{P} & \textbf{R} & \textbf{F1} & \textbf{Top-1} & \textbf{P} & \textbf{R} & \textbf{F1} & \textbf{Top-1}\\
\midrule
\textbf{\tool}     & 37.6 & \textbf{13.2} & \textbf{19.5} & \textbf{24.5} & 29.9 & \textbf{10.5} & \textbf{15.5} & \textbf{35.8}\\
\midrule
\textbf{\toolc}  & 28.7 & 12.7 & 17.6 & 22.3 & 25.2 & 10.0 & 14.3 & 30.2 \\ %
\textbf{\toola}  & \textbf{40.2} & 9.3 & 15.0 & 21.1 & \textbf{47.0} & 3.4 & 6.3 & 22.1 \\ %
\textbf{\toolv}  & 29.0 & 13.0 & 18.0 & 22.7 & 22.5 & 10.4 & 14.2 & 32.0 \\ %
\bottomrule[1pt]
\end{tabular}
\label{table:abalation_study}
\end{table}

The results of the ablation study are summarized in \autoref{table:abalation_study}. As expected, each variant of \tool demonstrated varying degrees of effectiveness compared to the full model. \toolc, which modifies the code line highlighting approach, showed a moderate reduction in performance, with the Top-1 accuracy dropping by 5\% compared to the full \tool, suggesting that the original highlighting strategy plays a critical role in precise vulnerability localization. In \toolv, which replaces contextual information with an MLP classifier, the accuracy dropped by 4\%, indicating that contextual information plays a certain role in aiding vulnerability classification. However, \toola, where attention outputs were compressed through average pooling, exhibited the most significant performance decline, with an F1 score reduced by 9\%. This underscores the importance of effective dimensionality reduction and feature selection techniques for accurate vulnerability localization, as not all parts of the attention output can capture the semantic meaning relevant to identifying vulnerabilities.

\vspace{2mm}

 \begin{mdframed}[backgroundcolor=red!2] 
 
\noindent\textit{\textbf{Answer (RQ4).} {The ablation study reveals that each component is essential for \tool’s performance. \toolc shows a moderate decline, while \toolv overlooks the importance of contextual information in vulnerability detection, resulting in reduced accuracy. \toola performs the worst due to the absence of appropriate dimensionality reduction and feature selection methods. }}
\end{mdframed}

\section{Related Work}

\noindent\textbf{Vulnerability Localization.}
Automated vulnerability localization can significantly reduce developers' efforts in identifying and diagnosing vulnerabilities.
Traditional vulnerability localization tools, such as~\cite{abreuAccuracySpectrumbasedFault2007, moonAskMutantsMutating2014}, rely on predefined detectors for different vulnerabilities, limiting scalability.
Deep learning-based methods~\cite{li2021vulnerability, li2021vuldeelocator, zhang2023learning, fu2022vulrepair, fu2022linevul, rajput2023icspatch, mirsky2023vulchecker,visalli2019towards} focus on learning different program features to determine whether a specific line contains a vulnerability. Program features include syntactic features (e.g., Abstract Syntax Tree~\cite{li2021vulnerability, li2021vuldeelocator}), semantic features (e.g., Program Dependency Graph~\cite{li2021vulnerability, mirsky2023vulchecker}), and other program representations. However, deep learning methods suffer from poor generalization. On the one hand, these methods are often limited to a specific programming language, and on the other hand, collecting various program features requires the program to be compilable or executable to gather relevant information, imposing additional requirements on the program structure.

With the rise of LLMs, LLM-based vulnerability localization has emerged~\cite{yao2024survey}. The key advantage of LLM-based approaches is their generalizability, allowing them to be applied across different types and domains of vulnerabilities. Current LLM-based methods can be broadly categorized into two types: direct output methods~\cite{nong2024chain, zhang2024empirical} and fine-tuning-based methods~\cite{yang2024large, zhang2024empirical}. Our approach innovatively leverages self-attention mechanisms as indicators for vulnerability localization, significantly improving accuracy compared to direct output by LLMs, without the high cost of fine-tuning. Additionally, \tool mitigates the performance degradation of LLM in long-context scenarios, making it better suited to real-world vulnerability localization needs.

\vspace{2mm}
\noindent\textbf{LLM Attention  for Code Analysis.} 
Self-attention mechanisms are widely used as interpretability tools for code. For code syntax interpretability, Wan et al.~\cite{wanWhatTheyCapture2022} used attention to assess whether pre-trained language models can capture syntactical relationships in source code. They observed that tokens sharing a common parent node in the Abstract Syntax Tree (AST) receive higher attention, suggesting that attention can help explain the model's ability to capture syntactic structures.   In terms of code semantic interpretability, Ma et al.~\cite{maUnveilingCodePreTrained2024} explored whether tokens within the same Control Dependency Graph (CDG), Control Flow Graph (CFG), or Data Dependency Graph (DDG) receive significantly increased attention. Their results indicate that many attention heads are indeed capable of capturing semantic information.  
Additionally, self-attention mechanisms can represent the importance of each token in a prompt relative to the output. Kou~\cite{kouLargeLanguageModels2024a} examined whether the content that LLMs focus on during coding tasks aligns with humans.  They used the last token in the prompt and the last token in the output as query tokens, calculating the attention to each token in the prompt as a measure of the LLM's focus during generation. We are the first to apply attention as an indicator for vulnerability localization. \looseness=-1

\section{Conclusion}

 Our work demonstrates that leveraging self-attention mechanisms within LLMs offers a powerful new approach to vulnerability localization. By systematically tracking attention weights, we can effectively identify lines of code that are more likely to contain vulnerabilities, even in large and complex codebases.  Our experiments show significant improvements in vulnerability localization accuracy, demonstrating the potential of self-attention-based approaches to advance the state of the art in software security. We hope our framework will open new possibilities for automating vulnerability detection, and reduce both the time and resources required for secure software development.

\bibliographystyle{ACM-Reference-Format}
\bibliography{ref}

\end{document}